\documentclass[aps,prl,twocolumn,nopacs,superscriptaddress]{revtex4-1} 

\usepackage{amsmath}
\usepackage{bbm}
\usepackage{epsfig}
\usepackage{epstopdf}
\usepackage[normalem]{ulem}
\usepackage{cancel}
\usepackage{mathtools}
\usepackage{paralist}

\usepackage{hyperref}

\newcommand{\id}{\mathbbm{1}}

\usepackage{times}

\newcommand{\nn}{{\mathbbm{N}}}
\newcommand{\zz}{{\mathbbm{Z}}}
\newcommand{\qq}{{\mathbbm{Q}}}

\DeclareMathOperator{\diag}{diag}
\DeclareMathOperator{\tr}{tr}

\usepackage{amsthm}

\begin{document}

\newtheorem{theorem}{Theorem}
\newtheorem{proposition}[theorem]{Proposition}
\newtheorem{lemma}[theorem]{Lemma}
\newtheorem{corollary}[theorem]{Corollary}
\newtheorem*{definition}{Definition}
\newtheorem{observation}[theorem]{Observation}
\newtheorem{example}[theorem]{Example}
\newtheorem{assumption}[theorem]{Assumption}

\title{Quantum measurement occurrence is undecidable}

\author{J.\ Eisert}
\affiliation{Qmio Group, Dahlem Center for Complex Quantum Systems, Freie Universit{\"a}t Berlin, 14195 Berlin, Germany}
\author{M.\ P.\ M{\"u}ller} 
\affiliation{Perimeter Institute for Theoretical Physics, 31 Caroline Street North, Waterloo, Ontario N2L 2Y5, Canada}
\author{C.\ Gogolin}
\affiliation{Qmio Group, 
Dahlem Center for Complex Quantum Systems, Freie Universit{\"a}t Berlin, 14195 Berlin, Germany}

\begin{abstract}
In this work, we show that very natural, apparently simple problems in quantum measurement theory can be undecidable even if their classical analogues are decidable.
Undecidability hence appears as a genuine quantum property here.
Formally, an undecidable problem is a decision problem for which one cannot construct a single algorithm that will always provide a correct answer in finite time. 
The problem we consider is to determine whether sequentially used identical Stern-Gerlach-type measurement devices, giving rise to a tree of possible outcomes, have outcomes that never occur.
Finally, we point out implications for measurement-based quantum computing and studies of quantum many-body models and suggest that a plethora of problems may indeed be undecidable.
\end{abstract}

\maketitle

At the heart of the field of quantum information theory is the insight that the computational complexity of similar tasks in quantum and classical settings may be crucially different.
Here we present an extreme example of this phenomenon: an operationally defined problem that is undecidable in the quantum setting but decidable in an even slightly more general classical analog. While the early focus in the field was on the assessment of tasks of quantum information processing, it has become increasingly clear that studies in computational complexity are also very fruitful when approaching problems outside the realm of actual information processing, for example in the field of Hamiltonian complexity \cite{Kempe,Terhal,Dorit,Osborne,Area}, or dynamical problems in channel theory~\cite{ChannelNPHard}.
In the meantime, a plethora of computationally hard tasks has been identified, both as far as \emph{NP-hard problems}
are concerned as well as their ``quantum analogues,'' the \emph{QMA-hard} ones.
These results show that it is presumably difficult to find an answer to those problems, but with sufficient computational effort, it can still be done.

Surprisingly, as will become clear, very natural decision problems in quantum theory may not only be computationally hard, but in fact even provably \emph{undecidable} \cite{FrustratedRemark,CubittPGWolf}, i.e., there cannot be an algorithm, or for that matter a standard Turing machine, that always provides the correct answer in finite time.
As such, this class of problems is in the same category as the halting problem that was famously shown to be undecidable in Alan Turing's work from 1936 \cite{Turing}.
The problem is to determine, given some program and an input, whether this program will eventually come to an end with that input -- so will ``halt'' -- or whether the program will continue running forever.
The key insight of Alan Turing was to recognize that there cannot be a single algorithm that is able to correctly answer every instance of that problem.
Of course, one can execute any algorithm for any finite time, but in case the program has then still not halted, one cannot judge in general whether or not it will ever do so.
This seminal insight has had profound implications in the theory of computing and in fact even to mathematics:
It implies G{\"o}del's first incompleteness theorem \cite{Goedel}, which states that a consistent, complete, and sound axiomatization of all statements about natural numbers cannot be achieved. 

In this work, we demonstrate that the very natural physical problem of determining whether 
certain outcome sequences cannot occur in repeated quantum measurements is undecidable, even though the same problem for classical measurements is readily decidable.
We do so by employing a reduction:
We show that if the problem that we introduce could always be solved, then one could find an algorithm that solved every instance of the 
halting problem -- which cannot be true.
At the same time we prove that the arguably most general classical analogue of the problem is always decidable, which shows that the undecidability is remarkably a genuine quantum mechanical feature. 

We also suggest that it is reasonable to expect a number of further such results, in particular in the context of quantum information and quantum many-body theory.

\begin{figure}[t]
\includegraphics[width=.8\columnwidth]{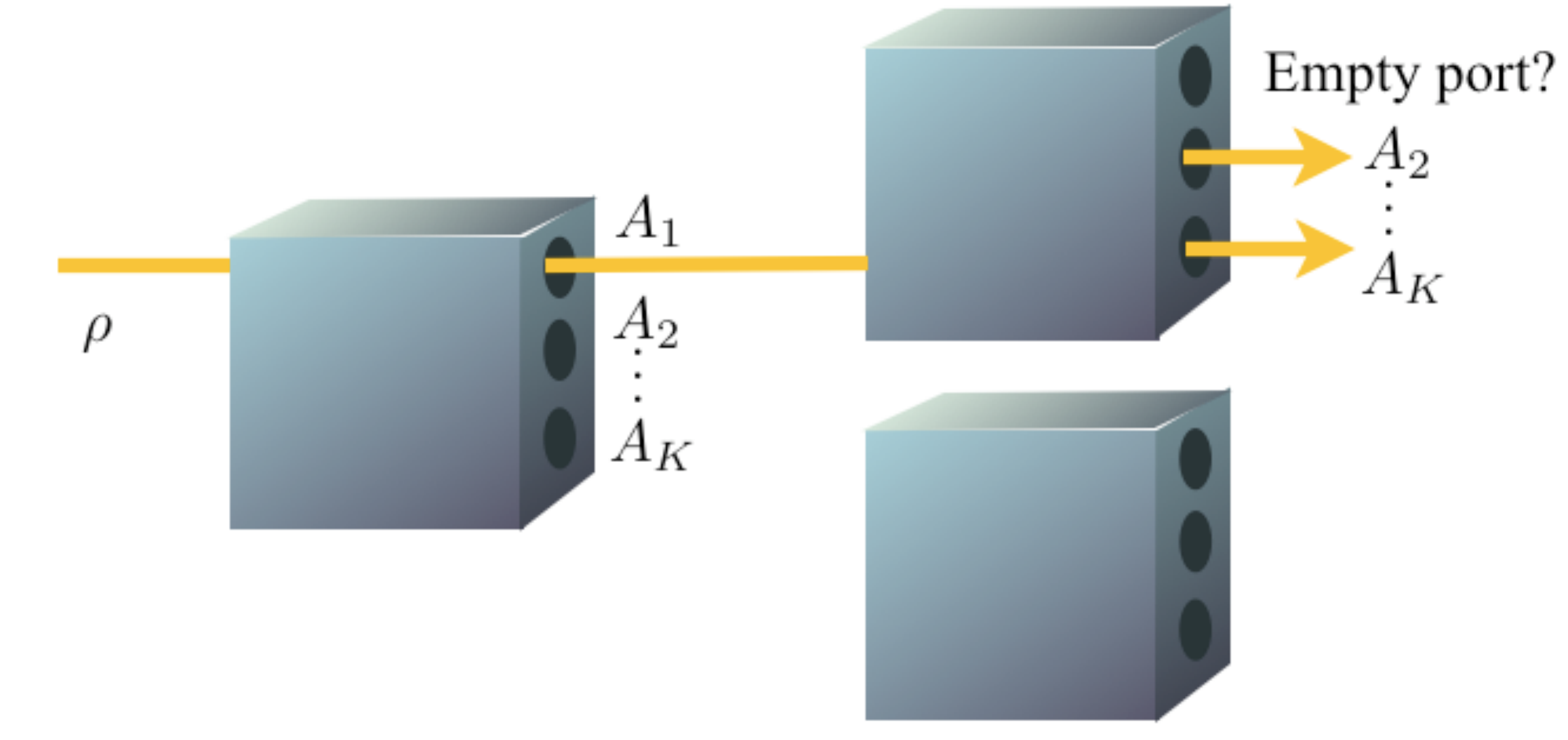}
\caption{
(Color online) 
The setting of sequential application of Stern-Gerlach-type devices considered here, gives rise to a tree of possible outcomes. The problem is to decide whether 
there exists an \emph{empty port} through which the particle will never fly.}
\label{fig_und}
\end{figure}

\paragraph{Setting.}
The decision problem that we will prove undecidable is motivated by the following natural quantum measurement setting:
Consider a measurement device that selectively measures a $d$-dimensional quantum system and has $K$ possible outcomes.
Such a device is a generalization of a \emph{Stern-Gerlach type} device that performs a nonprojective measurement.
The $K$ outcomes of the device are associated with Kraus operators $A_1,\dots, A_K$.
A measurement leading to outcome $j\in\{1,\dots, K\}$ occurs with probability $\tr(A_j \rho A_j^\dagger)$ and changes the state of the system according to
\begin{equation}
  \rho\mapsto \frac{A_j \rho A_j^\dagger}{\tr(A_j \rho A_j^\dagger)}.
\end{equation}	
In a sequence of $n$ measurements, the output state of such a device is repeatedly fed into an identical measurement device, leading to a tree of measurements (see Fig.~\ref{fig_und}).
Each path through this tree is associated with a sequence of outcomes $j_1,\ldots,j_n$.
In order to have a meaningful decision problem, where each input can be described by finitely many bits,
we restrict the problem to measurements whose Kraus operators are matrices of rational numbers:

\begin{definition}[Quantum measurement occurrence problem (QMOP)]
  Given a description of a quantum 
  measurement device in terms of $K$ Kraus operators
  $\{A_1,\dots, A_K\}\subset \qq^{d\times d}$, the
  task is to decide whether, in the setting described above,
  there exists any finite sequence of outcomes $j_1,\ldots,j_n$ 
  that can never be observed, even if the input state has full rank.
\end{definition}
Note that the notion of undecidability itself is independent of the physical theory: if a problem
is undecidable for classical Turing machines, then it is also undecidable for quantum Turing machines, and vice versa since they can mutually simulate each other.
Thus, our result says that the QMOP cannot be decided, no matter what physical resources we use to try to come to a decision.

Furthermore, note that in the QMOP one is supplied a perfect classical description of the quantum measurement device, 
and there is no ``quantum uncertainty'' in the description itself. Yet, we will see below that
desctructive interference in the working of the quantum device, as encoded in the Kraus operators, renders the quantum measurement occurrence problem undecidable, in contrast to its classical counterpart.

\paragraph{Undecidability of the quantum problem.}
Figuratively speaking in the metaphor of the Stern-Gerlach device with its tree of outcomes,
the problem is to decide whether there exists an \emph{empty port} somewhere in the tree through which the particle will never fly.
Surprisingly this turns out to be undecidable:
 
\begin{theorem}[Undecidability of the quantum problem] 
  \label{thm:qmopisundecidable}
  The QMOP for $K=9$ and $d=15$ is undecidable.
\end{theorem}

This statement is a consequence of the undecidability of the so-called
\emph{matrix mortality problem (MMP)}:
\emph{Given some finite set of integer matrices $\{M_1,\ldots, M_k\}$, is there any finite matrix product $M_{j_1}\ldots M_{j_n}$ that equals the zero matrix?}
In other words, does the semigroup generated by $\{M_1,\ldots,M_k\}$ contain the zero matrix?
One can show that the MMP is undecidable by reducing it to the so-called post correspondence problem (PCP) \cite{PCP,PCPComment} (see the Appendix). More specifically:

\begin{theorem}[Undecidability of the MMP \cite{Paterson,Halava}]
  The MMP is undecidable for $3\times 3$ integer matrix semigroups generated by $8$ matrices.
\end{theorem}
That is to say, there cannot be an algorithm that takes the input 
$ \{M_1,\dots, M_8\}\subset \zz^{3\times 3}$
and computes in finite time whether or not there exists a sequence $j_1,\dots, j_n$ such that 
$
  M_{j_1}\dots M_{j_n}=0.
$
In fact, in a  variant of the argument, the above theorem is still valid for semigroups generated by $7$
integer $3\times 3$ matrices \cite{New}.
Whether the MMP is also undecidable in the case of $2\times 2$ matrices is still an open problem~\cite{Miller}.

Turning back to the quantum problem, in terms of the Kraus operators, the probability for obtaining a particular sequence
$\mathbf{j}=j_1,\dots ,j_n$ of outcomes $j_i\in \{1,\dots, K\}$ is 
\begin{equation}
  \label{eqpw}
  p_{\mathbf{j}}= \tr(A_{j_n}\dots A_{j_1}\rho A_{j_1}^\dagger \dots A_{j_n}^\dagger).
\end{equation}
Now $\tr(A_{j_1}^\dagger \dots A_{j_n}^\dagger A_{j_n}\dots A_{j_1}\rho )=0$ for
 a full rank quantum state $\rho$ if and only if
$A_{j_1}^\dagger \dots A_{j_n}^\dagger A_{j_n}\dots A_{j_1}=0$.
Since this is a positive operator, the latter equality is true if and only if all of its singular values
are zero, i.e., if and only if $A_{j_n}\dots A_{j_1}=0$.


Now we relate an instance 
of a MMP to a set of suitable Kraus operators
$\{A_1,\dots,A_9\} \subset \qq^{15\times 15}$. Our approach is to take an instance of the
MMP, to encode it in Kraus operators having rational entries, and to 
complete them such that they form a trace-preserving completely positive map.
The key point of the argument is that although we extend
the dimension of the Kraus operators, 
a zero matrix is still found in the product of Kraus operators
exactly if and only if the corresponding MMP contains a zero matrix in the 
semigroup. A slight detour in the argument is necessary as we wish to arrive at Kraus operators
with rational entries.

For a given instance 
	$\{M_1,\dots, M_8\} \subset \zz^{3\times 3}$
of the MMP, define
\begin{equation}
	T\coloneqq \sum_{j=1}^8 
	M^\dagger_j M_j.
\end{equation}
Using the three integer matrices 
$P_1\coloneqq\diag(-1,1,1)$, $P_2\coloneqq\diag(1,-1,1)$, $P_3\coloneqq\diag(1,1,-1)$, and for $j\in\{1,\dots, 8\}$ set
\begin{eqnarray}
	M_{8+j} &\coloneqq&M_j P_1,\label{1}\\
	M_{16+j} &\coloneqq&M_j P_2,\label{2}\\ 
	M_{24+j} &\coloneqq&M_j P_3\label{3}.
\end{eqnarray}
This gives 
\begin{equation}
	\sum_{j=1}^{32}
	M_j^\dagger M_j = 4 \diag(T_{1,1},T_{2,2},T_{3,3}).
\end{equation}
Define $c\in \nn$ as
$
c \coloneqq \left\lceil 2 \left( \max\{T_{1,1},T_{2,2},T_{3,3}\} \right)^{1/2} \right\rceil .
$
By virtue of \emph{Lagrange's four-square theorem} \cite{foursquaretheorem},
every natural number can be written as the sum of four integer squares.
Hence, there exist four diagonal matrices $M_{33},\dots, M_{36}$ such that 
$
	\sum_{j=1}^{36}
	M_j^\dagger M_j = c^2\, \id_3 .
$
We now set for $j=1,\dots, 8$,
\begin{equation}
	A_j\coloneqq \frac 4 {5c}\left[
	\begin{array}{c|c}
	M_j &\\
	M_{8+j} &\\
	M_{16+j} & 0_{15\times12}\\
	M_{24+j} &\\
	M_{32+j} &\\
	\end{array}
	\right]
\end{equation}
with $M_{37}, \ldots, M_{40}\coloneqq 0_3$ and
\begin{equation}
	A_{9}\coloneqq \left(\frac 3 5\id_3\right)\oplus \id_{12}.
\end{equation}	
The matrices $\{A_1,\dots,A_9\} \subset \qq^{15\times 15}$ satisfy
$\sum_{j=1}^{9} A_j^\dagger A_j=\id_{15}$, as a simple calculation shows,
and thus describe a quantum measurement device.

We are now in the position to reduce the quantum measurement occurrence problem to the problem of deciding
whether the given semigroup contains the zero matrix. If this is the case, i.e., if
there exists a sequence $\mathbf{j}$ for which $M_{j_n} \dots M_{j_1}=0$, $j_1,\ldots,j_n\in\{1,\dots,8\}$, then $A_{j_n}\ldots A_{j_1}$ has the zero matrix
as its upper-left $3\times 3$ block. 
Moreover, the whole upper triangular matrix (including the diagonal) is zero as
well, which means that the matrix is nilpotent: there is some $m\leq15$ such that
$
   (A_{j_n}\ldots A_{j_1})^m = 0.
$

Conversely, let us assume that 
there exists an outcome sequence that is never observed, 
so there exists a sequence $\mathbf{j}$ such that $A_{j_n} \dots A_{j_1}=0$. Let $\mathbf{v}$ be the sequence that is obtained from $\mathbf{j}$
by omitting all $j_i$ for which $j_i=9$. Then, by construction, $M_{v_{|v|}}\dots M_{v_1}=0$. Therefore, the semigroup generated by $\{M_1,\dots, M_8\}$ contains the zero matrix.
\qed
%
%

The QMOP as described so far asks whether certain outcome sequences have probability \emph{exactly} equal to zero. From a physical point of view,
it is interesting to note that this result is to some extent robust, in the sense that it remains valid if small nonzero probabilities are allowed.
To see this, write every Kraus operator in the form $A_j=Z_j/N_j$, where $N_j\in\nn$ and 
$Z_j\in\zz^{d\times d}$.
Then the probability of the sequence $\mathbf{j}$ from Eq.~\eqref{eqpw} becomes
\begin{eqnarray}
  p_{\mathbf{j}}&=&(N_{j_1}\ldots N_{j_n})^{-2}\tr(\rho Z)\geq \frac 1 d N^{-n} \tr (Z)\nonumber\\
  &\geq &(d N)^{-n}\tr (Z),
\end{eqnarray}
where $Z\in \zz^{d\times d}$ fulfils $Z\geq 0$ and $N:=\max_j N_j^2\in\nn$. Thus, $p_{\mathbf{j}}$ is either exactly zero or
not less than $\delta^n$, where $\delta:=1/(dN)$ is a function of the Kraus operators. Therefore, the QMOP is equivalent to the following problem:
\emph{Given $K$ Kraus operators $\{A_1,\ldots,A_K\}\subset\qq^{d\times d}$, is there a finite sequence of outcomes $j_1,\ldots,j_n$ which has probability less than $\delta^n$
(with $\delta>0$ defined above)
if the input is the maximally mixed state?}

\paragraph{Decidability of the classical problem.}

We now turn to a corresponding classical problem, the \emph{classical measurement occurrence problem} (CMOP).
A classical channel is described by a stochastic matrix $Q$ acting on $d$-dimensional probability vectors $\vec{q}$.
A description of a classical selective measurement device with $K$ outcomes, is given by a decomposition $Q=\sum_{j=1}^K Q_j$ into matrices $Q_1,\dots , Q_K$ with non-negative entries
(such matrices are sometimes called \emph{substochastic}), that specify the action of the device on the classical system.
That is, on outcome $j$ the probability vector is transformed according to
\begin{equation}
  \vec{q} \mapsto \frac{Q_j \vec{q}}{\sum_{i=1}^d (Q_j \vec{q})_i} .
\end{equation}
This is arguably the most general classical analog of the QMOP.
The probability for obtaining a particular sequence $j_1,\dots, j_n$
of outcomes $j_i\in \{1,\dots, K\}$ on an input probability vector $\vec{q}$ is $\sum_{i=1}^d (Q_{j_n}\ldots Q_{j_1} \vec{q})_i$.
This is zero for an input vector $\vec{q}$ with all $(\vec{q})_i>0$ if and only if $Q_{j_n}\ldots Q_{j_1}=0$.
The CMOP is thus obviously equivalent to the MMP with entrywise non-negative matrices.
For this case the MMP is decidable, which was shown in Ref.~\cite{Blondel} for $K=2$, and the general case follows by an essentially equivalent argument.

It shall be noted that our definition of classical devices is even more general than that of the quantum devices considered before;
it represents the most general form of any conceivable classical measurement device.
Namely, we allow for mixing in each outcome, which would in the quantum case correspond to a device that applies a whole quantum channel, not just a single Kraus operator, per outcome. 

We now turn to proving decidability of the MMP with elementwise non-negative Kraus operators from which decidability of the classical case and for a subclass of quantum measurement devices follows.

\begin{theorem}[Decidability of the non-negative MMP] \label{thm:decidabiliy}
The MMP is decidable for any $d\times d$ matrix semigroup generated by $K$ matrices with non-negative rational entries.
\end{theorem}

Although the MMP is decidable for matrices with non-negative entries, it is still a hard problem: even in the case of $K=2$ matrices, this problem is NP-complete~\cite{Blondel}.

\begin{corollary}[Decidability of the classical problem]
  For any $K$ and $d$, both the \emph{quantum measurement occurrence problem (QMOP)} with Kraus operators $\{A_1,\dots, A_K\} \subset {\qq_0^+}^{d\times d}$ with non-negative entries and the \emph{classical measurement occurrence problem (CMOP)} are decidable.
\end{corollary}

In order to prove Theorem~\ref{thm:decidabiliy} we introduce some notation first. For an elementwise non-negative matrix $M$ we define the matrix $M'$ elementwise by
\begin{align}
M'_{a,b}\coloneqq \begin{cases}
0 & \text{if } M_{a,b}=0,\\
1 & \text{if } M_{a,b}>0 .
\end{cases}
\end{align}
For two such binary matrices $M',N'$ we define their associative binary matrix product by
$M' * N'\coloneqq (M' N')'$.
Note that $M_{j_1} \ldots  M_{j_n} = 0$ if and only if $(M_{j_1} \ldots  M_{j_n})' = 0$, which in turn holds if and only if 
$
  M'_{j_1} * \ldots *  M'_{j_n} = 0.
$
As all matrices in the semigroup $\cal S$ generated by $S=\{M'_1, \ldots , M'_K\}$ under the matrix multiplication $\ast$ are binary matrices, hence 
$|{\cal S}| \leq 2^{(d^2)}$. We finish the proof by arguing that every element $M'$ of $\cal S$ can be written in terms of at most $|\cal S|$ elements from $S$. Fix some $M'$ and let $j_1,\ldots,j_m$ be the shortest sequence of indices such that $M'= M'_{j_m}\ast\ldots\ast M'_{j_1}$. Then for all $k < l \leq m$ we have $M'_{j_l}\ast \ldots \ast M'_{j_1} \neq M'_{j_k}\ast \ldots \ast M'_{j_1}$, because otherwise we would obtain a shorter representation of $M$ by replacing the former product with the latter.
Therefore, for each $l\leq m$ the product $M'_{j_l}\ast \ldots \ast M'_{j_1}$ yields a different elements of $\cal S$ and hence $m\leq |{\cal S}| \leq 2^{(d^2)}$.

\paragraph{Outlook and implications for quantum many-body problems.}
We have seen in this work that very natural decision problems in quantum measurement theory
can be undecidable, even if their classical counterparts are decidable.
In the specific problem that we considered (quantum measurement occurrence problem), the existence of negative transition matrix elements
renders the quantum problem more complex than its classical counterpart --
that is, the effect of \emph{destructive interference}.
We conclude by a number of further comments: 

Firstly, note that mild variants of the above problem can easily
lead to problems that have efficient solutions.
For example, if one considers trace-preserving quantum channels, one can give upper bounds to the number of times a channel must be applied, so that it maps any density operator to one with full rank, by virtue of the quantum Wielandt theorem \cite{Sanz}. Thus, the problem whether
there is some $n$ such that the $n$-fold application of a nonselective channel yields nonzero
probabilities, for all subsequent measurements and for all inputs, is
efficiently decidable.

Second, the above statement has immediate implications to undecidability
in quantum many-body physics \cite{MBP} and quantum computing.
Interpreting the above matrices $\{A_1,\dots, A_K \}$ as those defining matrix-product
states \cite{Area,Gross,MPS}, several other natural undecidable problems open up.

As an example, consider a family of one-dimensional quantum wires for \emph{measurement-based quantum computing}
in the sense of Refs.~\cite{Gross}.
These wires are described by families of 
matrix-product states of length $n$, being defined by products of matrices $\{A_1,\dots, A_K\}$
(the same set of matrices is taken for each site),
associated with measurement outcomes $1,\dots, K$ in the computational basis. The left
and right boundary conditions 
are fixed as $|L\rangle=|R\rangle= [1 \, 0 \,\dots 0]^T$. The task is to determine whether 
there exists a sequence of measurement outcomes $j_1,\dots, j_n$ that will never occur \cite{MPSFoot}.
The subsequent result is a consequence of the above reasoning, together with the fact that
the problem whether the semigroup generated by integer matrices contains a matrix
with a zero element in the left upper corner is undecidable \cite{New}.

\begin{theorem}[Undecidability in quantum computing]
  Given a description of a family of matrix-product states defined by the matrices
  $\{A_1,\dots, A_K\}\subset \qq^{d\times d}$,
  the  task is to decide whether there exists an $n$ and a sequence of outcomes $j_1,\ldots,j_n$ for a
  wire of length $n$ of
  local measurements in the computational basis
  that will never be observed. This problem is undecidable.
\end{theorem}
Similar reasoning as in the proof of the undecidability of the quantum measurement occurrence problem suggests that other questions concerning the characterization of measurement outcomes are undecidable as well.
These observations indicate that undecidability may be a natural and frequent phenomenon in many-body quantum physics and computation.

Similarly interestingly, a number of problems in quantum information theory seem to be natural candidates for being potentially undecidable. This applies notably to the problem of deciding whether a quantum state is distillable, giving a new perspective to the notorious question of deciding whether bound entangled states with a negative partial transposition exist.

\emph{Note added.} Compare also the recent related independent work Ref.~\cite{WolfCubittPerezGarcia}.

\paragraph{Acknowledgments.}
We would like to thank M.\ Kliesch for discussions and the
EU (QESSENCE, MINOS, COMPAS), the German National Academic Foundation,
the BMBF (QuOReP), the Government of Canada (Industry Canada), the
Province of Ontario (Ministry of Research and Innovation), 
and a EURYI for support.

\appendix

\section*{Appendix}

For the readers convenience we very briefly sketch the elements of the argument relating
the MMP to the PCP. We consider the PCP over the two alphabets $\Sigma$
and $\Delta$, where $\Sigma$ is arbitrary and $\Delta=\{2,3\}$. Even though $\Delta$ is fixed, this version of the PCP
is still undecidable \cite{Halava}.
In order to relate this problem to a matrix problem, set $\Gamma\coloneqq\{1,2,3\}$ and 
consider the map
$f:\Gamma^\ast \rightarrow \nn$ defined as
\begin{equation}
	f(w) = \sum_{j=1}^{|w|} w_j 3^{|w|-j}
\end{equation}
for all nonempty words $w$ over $\Gamma$, where $|w|$ denotes the length of $w$. 
$f(w)$ is the $3$-adic representation of $w$. 
Now continue to define the 
function $F: \Gamma^\ast \times \Gamma^\ast\rightarrow \nn^{3\times 3}$ as
\begin{equation}
	F(u,v) = \left[
	\begin{array}{ccc}
	1 & 0 & 1\\
	1 & 1 & 0\\
	0 & 0 & 1\\
	\end{array}
	\right]
	\left[
	\begin{array}{ccc}
	3^{|u|} & 0 & 0\\
	0 & 3^{|v|} & 0\\
	f(u) & f(v) & 1\\
	\end{array}
	\right]
	\left[
	\begin{array}{ccc}
	1 & 0 & 1\\
	1 & 1 & 0\\
	0 & 0 & 1\\
	\end{array}
	\right]^{-1}.
\end{equation}
Let now $(h,g)$ be an instance of PCP, $h,g:\Sigma^\ast\rightarrow \Delta^\ast$.
For each such 
instance, define the $3\times 3$-matrices
\begin{equation}
	X_a= F(h(a),g(a)),\,
	Y_a= F(h(a),1 g(a))
\end{equation}
for $a\in \Sigma$. Let $S$ be the matrix semigroup generated by $\{X_w,Y_w: w\in \Sigma\}$. 
One then continues to consider matrix products 
\begin{equation}
	M= M_{w_1}\dots M_{w_n}\in S 
	\label{eqProduct}
\end{equation}
for a given word 
$w=w_1\dots w_n$, where $M_{w_j}= X_{w_j}$ or $M_{w_j}= Y_{w_j}$.
The key step of the proof of Ref.~\cite{Halava}, deriving from the encoding of 
Ref.~\cite{Paterson}, is to show that $M_{1,1}=0$ (denoting the upper left
element of the matrix) holds true
if and only if $w$ is a solution of the instance $(h,g)$. This shows that the problem to
decide whether the semigroup contains an element the upper left element of which 
is zero is undecidable. 
By adding the idempotent matrix
\begin{equation}
	B= \left[
	\begin{array}{ccc}
	1 & 0 & 0\\
	0 & 0 & 0\\
	0 & 0 & 0
	\end{array}
	\right]
\end{equation}
as an additional generator to the set of matrices $\{X_w,Y_w: w\in \Sigma\}$,
it is then a simple step to reduce the MMP to the PCP.
In Ref.~\cite{Halava}, it is shown that we may choose $|\Sigma|=7$ and specific forms of the
product \eqref{eqProduct}, which gives a count of exactly $8$ matrix generators.


\begin{thebibliography}{99}
\bibitem{PCPComment}
	Ref.~\cite{BlondelQCA} relates the problem whether the 
	language recognized by a quantum finite automaton is empty or nonempty to the PCP.
\bibitem{MPSFoot}
	Note that for a matrix-product state vector $|\psi\rangle$ of the above
	kind, $\langle j_1,\dots, j_n|\psi\rangle = \langle R| A_{j_n}\dots A_{j_1}|L\rangle$.
\end{thebibliography}
\end{document}